%% file: main.tex
\documentclass[sigconf]{acmart}
\renewcommand\footnotetextcopyrightpermission[1]{}  

\setcopyright{none}

\AtBeginDocument{%
  }
\usepackage{algorithmic}
\usepackage{graphicx}
\usepackage{textcomp}

\usepackage{enumitem} 
\usepackage{amsmath,amsthm,amsfonts} 

\usepackage{amssymb}                   

\usepackage{mathtools}                 
\usepackage{dsfont}                    
\usepackage[table,xcdraw]{xcolor}      
\usepackage{multirow}
\usepackage{tikz}

\usepackage{url}
\usepackage{xspace} 

\newcommand{\sysname}{\textsf{FHE-Agent}\xspace}

\begin{document}

\title{FHE-Agent: Automating CKKS Configuration for Practical Encrypted Inference via an LLM-Guided Agentic Framework}


\author{Nuo Xu}
\affiliation{
  \institution{University of Minnesota -- Twin Cities}
  \country{}
}
\email{xu001536@umn.edu}

\author{Zhaoting Gong}
\affiliation{
  \institution{North Carolina State University}
  \country{}
}
\email{zgong6@ncsu.edu}

\author{Ran Ran}
\affiliation{
  \institution{North Carolina State University}
  \country{}
}
\email{rran@ncsu.edu}

\author{Jinwei Tang}
\affiliation{
  \institution{University of Minnesota -- Twin Cities}
  \country{}
}
\email{tang0940@umn.edu}

\author{Wujie Wen}
\affiliation{
  \institution{North Carolina State University}
  \country{}
}
\email{wwen2@ncsu.edu}

\author{Caiwen Ding}
\affiliation{
  \institution{University of Minnesota -- Twin Cities}
  \country{}
}
\email{dingc@umn.edu}


\input{Sections/00_abstract}

\maketitle

\input{Sections/01_introduction}

\input{Sections/02_related_work}

\input{Sections/03_threat_model}
\input{Sections/04_methodology}
\input{Sections/05_evaluation}
\input{Sections/06_conclusion}

\clearpage
\bibliographystyle{ACM-Reference-Format}

\bibliography{reference.bib}

\end{document}

%% file: Sections/00_abstract.tex
\begin{abstract}
Fully Homomorphic Encryption (FHE), particularly the CKKS scheme, is a promising enabler for privacy-preserving MLaaS, but its practical deployment faces a prohibitive barrier: it heavily relies on domain expertise. Configuring CKKS involves a tightly coupled space of ring dimensions, modulus chains, and packing layouts. Without deep cryptographic knowledge to navigate these interactions, practitioners are restricted to compilers that rely on fixed heuristics. These "one-shot" tools often emit rigid configurations that are either severely over-provisioned in latency or fail to find a feasible solution entirely for deeper networks.

We present FHE-Agent, an agentic framework that automates this expert reasoning process. By coupling a Large Language Model (LLM) controller with a deterministic tool suite, FHE-Agent decomposes the search into global parameter selection and layer-wise bottleneck repair. The agents operate within a multi-fidelity workflow, pruning invalid regimes using cheap static analysis and reserving expensive encrypted evaluations for the most promising candidates.

We instantiate FHE-Agent on the Orion compiler and evaluate it on standard benchmarks (MLP, LeNet, LoLa) and deeper architectures (AlexNet). FHE-Agent consistently achieves better precision and lower latency than naïve search strategies. Crucially, it automatically discovers feasible, 128-bit secure configurations for complex models where baseline heuristics and one-shot prompts fail to produce a valid setup.

\end{abstract}

%% file: Sections/01_introduction.tex
\section{Introduction}
Machine learning as a service (MLaaS) has become the dominant paradigm for deploying deep neural networks, yet it sharply amplifies privacy concerns: sending raw inputs to third-party infrastructure risks leaking sensitive attributes~\cite{hesamifard2018ppml}.
Fully homomorphic encryption (FHE) offers a principled cryptographic remedy: it allows a cloud server to compute directly on encrypted data and return encrypted predictions without ever accessing the underlying plaintext~\cite{hesamifard2018ppml,cheon2017ckks,guardml2024}.
For deep learning, approximate schemes such as CKKS~\cite{cheon2017ckks} enable end-to-end encrypted inference by supporting real-valued arithmetic and SIMD-style packing.

However, this capability comes with a steep performance tax.
Even with carefully engineered circuits, FHE operations are typically $10^3$–$10^5\times$ slower than their cleartext counterparts~\cite{ran2023spencnn,krastev2024fhelipe,cryptoracle2025}.
Unoptimized CPU implementations of ResNet-20 on CIFAR-10 can require thousands of seconds per encrypted image~\cite{lee2022resnet20}, and state-of-the-art CPU compilers like Orion still report hundreds of seconds for moderately sized networks~\cite{ebel2025orion}.
Recent GPU accelerators like Cheddar reduce this to a few seconds per image~\cite{kim2024cheddar}, but performance is still dominated by the underlying FHE configuration: a poorly chosen modulus chain, packing layout, or bootstrapping schedule can easily turn a viable system into one that times out or fails to decrypt.

Unfortunately, navigating this configuration space is notoriously difficult.
A practitioner must simultaneously tune
(i) the CKKS parameter set $(\log N,\log Q)$ and scale schedule;
(ii) the bootstrapping plan and depth budget; and
(iii) layer-specific packing layouts and batch sizes.
These decisions are tightly coupled and highly non-linear:
aggressive packing improves throughput but explodes rotation costs;
conservative scaling preserves precision but inflates the modulus chain;
and a chain that is too short for the network’s depth makes encrypted inference impossible.
While compilers such as CHET, Orion, and Fhelipe~\cite{dathathri2019chet,ebel2025orion,krastev2024fhelipe} raise the abstraction level, they typically generate \emph{one} functional configuration per model using fixed heuristics.
Even optimizers like AutoPrivacy~\cite{lou2020autoprivacy} focus on restricted settings and do not target the full design space of CKKS-based deep networks.

Empirically, deploying a new model remains a manual, error-prone search: start from a reference, run encrypted inference, diagnose failures or timeouts, tweak parameters, and repeat~\cite{krastev2024fhelipe,cabreroThesis2023}.
Yet modern FHE backends already expose rich internal signals: static analyzers estimate security; logical simulators track noise and slot utilization; and profilers highlight bottlenecks~\cite{ebel2025orion,cryptoracle2025,ran2023spencnn}.
The barrier is no longer \emph{observability}, but rather how to turn these signals into an automated, resource-aware search over configurations.

These observations expose three practical challenges:
\begin{itemize}[leftmargin=*]
\item \textbf{(C1) Cross-layer, multi-objective coupling.} Configuration choices interact in subtle, model-specific ways; changing a global knob (e.g., the scale schedule) can break precision or security several layers away, making latency/accuracy/security trade-offs hard for non-experts.
\item \textbf{(C2) Scarce encrypted-evaluation budget.} Full FHE runs cost seconds to minutes even on accelerators~\cite{kim2024cheddar}, so naively exploring many configurations with encrypted inference is infeasible.
\item \textbf{(C3) Under-utilized backend signals.} Existing compilers treat encrypted inference as a mostly monolithic step; they do not expose a first-class, multi-fidelity workflow where static analysis and cleartext simulation aggressively prune the search space before execution.
\end{itemize}

Large language models (LLMs) acting as \emph{agents} provide a promising substrate to address these challenges.
Frameworks like AutoGen and Reflexion~\cite{wu2023autogen,shinn2023reflexion} show that LLM agents can plan and refine designs from feedback.
In the FHE domain, TFHE-Coder~\cite{kumar2025tfhe} uses agents to synthesize secure Boolean circuits, but targets functional correctness rather than performance optimization for CKKS.
To our knowledge, no system leverages an LLM controller to orchestrate CKKS compilers, explicitly treat encrypted evaluation as a scarce resource, and navigate the joint configuration space of scheme parameters, packing, and bootstrapping under user-level constraints.

\noindent\textbf{In this paper.}
We introduce \textbf{\sysname}, an agentic framework that automates the configuration
of CKKS-based encrypted inference.
Instead of treating the backend as a black box, \sysname factors an existing framework
(e.g., Orion) into a deterministic \emph{tool suite} exposing static analysis,
layerwise profiling, and bootstrapping constraints.
On top of this, a \emph{multi-agent controller} proposes discrete optimization
directions---from global CKKS configurations to local packing tweaks---and executes
them inside a \emph{multi-fidelity workflow} that aggressively exploits cheap static
and cleartext signals while treating encrypted evaluations as a scarce resource.
We prototype \sysname on Orion+Lattigo and evaluate it on MLPs, LeNet, LoLA, and
AlexNet, where it consistently finds feasible 128-bit–secure configurations and
reduces encrypted inference latency compared to naïve FHE sweeps and one-shot
LLM suggestions.

Concretely, we make the following contributions:
\begin{itemize}[leftmargin=*]
  \item We articulate the gap between one-shot FHE compilation and practical deployment
        as a \emph{resource-constrained configuration search} problem, highlighting
        challenges C1–C3.
  \item We design \sysname, which (i) decomposes an FHE configuration into global CKKS
        parameters and layer-local overrides, (ii) factors a backend into a reusable
        tool suite of analyzers, profilers, and cost models, and (iii) uses a hierarchical
        multi-agent controller to navigate this structured space using safe, interpretable
        directions.
  \item We instantiate a three-phase, multi-fidelity workflow: Phase~A explores structural
        regimes via static checks and cleartext simulation; Phase~B calibrates a latency
        cost model with sparse encrypted runs; and Phase~C performs admission-controlled
        refinement around a calibrated baseline under a strict encrypted-evaluation budget.
\end{itemize}

%% file: Sections/02_related_work.tex
\section{Background and Related Work}
\label{sec:background}

\subsection{FHE Encrypted Inference using CKKS}

CKKS encodes a real-valued vector into a polynomial in $\mathbb{C}[X]$ $/(X^N{+}1)$, enabling SIMD-style computation with up to $N/2$ complex slots per ciphertext. The parameter set ring degree $N$, ciphertext modulus $Q$, and scale $\Delta$ jointly determines slot capacity, numerical precision, and multiplicative depth. Homomorphic operations include plaintext--ciphertext multiplication (\texttt{PMult}), ciphertext--ciphertext multiplication (\texttt{CMult}), additions (\texttt{PAdd}, \texttt{CAdd}), relinearization, rescaling, and slot rotations. Since each \texttt{CMult} consumes one modulus level, insufficient remaining depth triggers bootstrapping~\cite{bossuat_efficient_2021, chen_improved_2018}, which itself requires multiple internal multiplications and rotations. Consequently, practical encrypted inference typically adopts $\log N \ge 16$ to support model depth plus at least one bootstrap.

In semi-honest inference settings, model parameters remain in plaintext while client inputs are encrypted. Linear and convolution layers decompose into sequences of \texttt{PMult}, \texttt{Rotate}, and \texttt{CAdd} that implement homomorphic MACs~\cite{juvekar_gazelle_2018, ebel2025orion, aharoni_helayers_2023}. Nonlinearities are approximated by low-degree polynomials~\cite{chen_-x_2022, park_powerformer_2024, dowlin_cryptonets_nodate}, and pooling is realized via pre-encoded binary masks~\cite{lu_bumblebee_2023, krastev2024fhelipe}. These patterns imply that rotations, plaintext multiplications, and polynomial evaluation dominate runtime, and they directly shape packing strategy: how activations are arranged across slots, how weights are encoded, and how many rotations each homomorphic MAC requires. The interplay between packing geometry, supported operations, and noise management underlies the design of efficient encrypted inference.

\subsection{FHE Libraries and Compilers}

\textbf{General-purpose libraries.}
Low-level CKKS libraries such as SEAL~\cite{laine2017seal}, OpenFHE~\cite{albadawi2022openfhe}, and Lattigo~\cite{lattigo} expose core primitives (e.g., \texttt{CMult}, rotations, key-switching) and basic safety checks, but leave parameter selection, noise budgeting, and packing strategy to developers. They typically serve as backend executors for higher-level systems.

\textbf{Frontend DSLs and compilers.}
To raise the abstraction level, systems such as CHET~\cite{dathathri2019chet}, HECO~\cite{haller2023heco}, Porcupine~\cite{cowan_porcupine_2021}, FHELipe~\cite{krastev2024fhelipe}, and Orion~\cite{ebel2025orion} compile linear algebra or small ML models into HE circuits, automating basic noise management and CKKS parameter selection. However, these tools generally target narrow model families and operate in a one-shot manner, producing a single configuration with limited ability to explore broader design spaces or user constraints.

\textbf{Middle-end optimization layers.}
Frameworks like EVA~\cite{dathathri_eva_2020} provide algebraic simplifications, but CKKS cost remains driven by noise growth and multiplicative depth. Optimizing scale management, depth allocation, and bootstrapping placement has been framed as a constrained optimization problem, motivating analytical and search-based approaches~\cite{liu_resbm_2025, cheon_halo_2025, lee_elasm_2023}. Systems such as AutoPrivacy~\cite{lou2020autoprivacy} and Cabrero-Holgueras et al.~\cite{cabrero2023params} explore these trade-offs via RL or rule-based search, but treat the HE backend as a black box and offer limited per-layer diagnostics.

\textbf{Limitations.}
Two limitations persist across these lines of work.  
(1) Rich domain knowledge, such as layer-wise depth propagation, rotation costs, and slot utilization, are often embedded inside compiler passes and rarely exposed to users or available for external tooling. Developers lack programmatic access to per-layer profiles, bottlenecks, or bootstrapping schedules.  
(2) Existing systems do not view FHE evaluation as a scarce resource. They offer limited support for multi-fidelity workflows that combine static analysis, cleartext simulation, and selective encrypted execution to prune large configuration spaces efficiently.

\textbf{Our approach.}
In contrast, \sysname\, decomposes an existing backend (e.g., Orion) into explicit, modular tools and introduces an agentic controller that systematically searches over FHE configurations under a strict encrypted-evaluation budget. This design exposes the necessary diagnostics for informed decision-making while enabling efficient multi-fidelity exploration of packing, scheduling, and parameter choices.

\subsection{LLM-Based Agentic Workflows}

Large language models (LLMs) are increasingly deployed as \emph{agents} that plan, invoke tools, and self-correct.
Frameworks like AutoGen~\cite{wu2023autogen} enable multiple agents to coordinate on complex tasks via conversation, invoking external tools such as compilers and debuggers in the loop.
Reflexion~\cite{shinn2023reflexion} and Voyager~\cite{wang2023voyager} demonstrate that LLM agents can improve through self-critique and acquire skills in open-ended environments by writing and reusing code.

The first work to combine LLM agents with HE is TFHE-Coder~\cite{kumar2025tfhe}, which synthesizes Fully Homomorphic Encryption Scheme Over the Torus (TFHE) programs using multi-agent verification.
However, TFHE-Coder targets functional correctness and security for bit-level circuits, not performance optimization for approximate schemes like CKKS.
To our knowledge, no prior system uses an LLM controller to orchestrate CKKS compilers and simulators as tools, in order to jointly optimize CKKS parameter sets, packing, and bootstrapping under accuracy, security, and latency constraints.
\sysname fills this gap by exposing the FHE backend as a toolbox of analyzers, profilers, and evaluators, and by using agents to refine FHE configurations within a multi-fidelity loop.

%% file: Sections/03_threat_model.tex
\section{Threat Model and Deployment Setting}
\label{sec:problem}
\subsection{Threat Model}
We consider a semi-honest cloud server that hosts the plaintext model and runs the FHE backend together with \sysname.
Before serving real client queries, the server enters an offline optimization phase in which an LLM-based agent generates the encrypted-inference code, selects FHE configurations, and verifies functional correctness on public calibration data or encrypted dummy inputs supplied by the client.
In this phase, the agent and tools observe only model weights, tool outputs, and aggregate metrics (e.g., depth usage, noise margins, timing), and never see any client plaintext.

Once a configuration has been selected, the server sends the client a description of the required tensor-to-slot layout and the public parameters and evaluation keys needed for inference.
The client applies this layout to its private inputs, encrypts them under its CKKS public key, and uploads only ciphertexts and evaluation keys.
The cloud server then executes the agent-generated inference code on these ciphertexts and returns encrypted outputs, which the client decrypts locally.

We assume the server is honest-but-curious: it follows the protocol but may attempt to infer information from observed ciphertexts and metadata.
We target standard lattice-based IND-CPA security at a target level (e.g., 128 bits) and do not address side-channel or traffic-analysis attacks.

\subsection{Config Space and Optimization Problem}
\label{subsec:problem_formulation}

Let $f_{\theta}$ be a trained neural network and $\mathcal{D}_{\mathrm{val}}$ a validation dataset.
An FHE configuration is a tuple
\begin{equation}
\footnotesize
  \label{eq:config_tuple}
  \mathcal{C}
  = (\log N, \log Q, \text{scale schedule}, \text{bootstrapping plan}, \text{packing scheme}),
\end{equation}
which fully determines the encrypted inference pipeline $\widetilde{f}_{\theta,\mathcal{C}}$ produced by the backend.
The pair $(\log N, \log Q)$ here corresponds to the CKKS parameter set; the remaining fields specify the scale schedule, bootstrapping decisions, and packing/layout choices.
We structure the search space by viewing $\mathcal{C}$ as a hierarchical object:
\begin{equation}
  \label{eq:hier_config_again}
  \mathcal{C}
  = \big(\mathcal{C}_{\mathrm{global}}, \{\mathcal{C}^{(i)}_{\mathrm{local}}\}_i\big),
\end{equation}
where $\mathcal{C}_{\mathrm{global}}$ comprises scheme-level parameters (including the CKKS parameter set) and global backend options, and each $\mathcal{C}^{(i)}_{\mathrm{local}}$ contains layer-specific packing overrides.
This decomposition aligns with our multi-agent controller's design (Section~\ref{sec:system}) and with how human experts reason about FHE deployments.

We treat the FHE backend as a deterministic oracle exposed via a unified API:
\begin{equation}
  \label{eq:run_trial_oracle}
  \texttt{run\_trial}(\mathcal{C}, \texttt{eval\_mode}) \rightarrow \texttt{metrics}.
\end{equation}
The \texttt{eval\_mode} selects the fidelity:
\texttt{STATIC\_ONLY} performs graph and parameter checks;
\texttt{CLEAR\_ONLY} runs floating-point simulation to profile precision, noise usage, and per-layer primitive counts;
and \texttt{FHE\_LIGHT}/\texttt{FHE\_FULL} execute actual encrypted inference on subsets or the full validation set.
These modes correspond directly to the tools in \sysname's multi-fidelity backend.

The optimization goal is to select an FHE configuration $\mathcal{C}$ that satisfies user-defined constraints---for example, accuracy within $\varepsilon$ of the plaintext model on $\mathcal{D}_{\mathrm{val}}$, at least 128-bit security, and latency below a target budget---while minimizing deployment cost.
\sysname addresses this by exploring the space defined by~\eqref{eq:hier_config_again} using the oracle in~\eqref{eq:run_trial_oracle}, strictly managing the budget of expensive encrypted evaluations through a three-phase, multi-fidelity workflow.

%% file: Sections/04_methodology.tex
\section{FHE-Agent System Design and Optimization Workflow}
\label{sec:system}

\begin{figure*}[t]  
    \centering  
    \includegraphics[width=.9\linewidth]{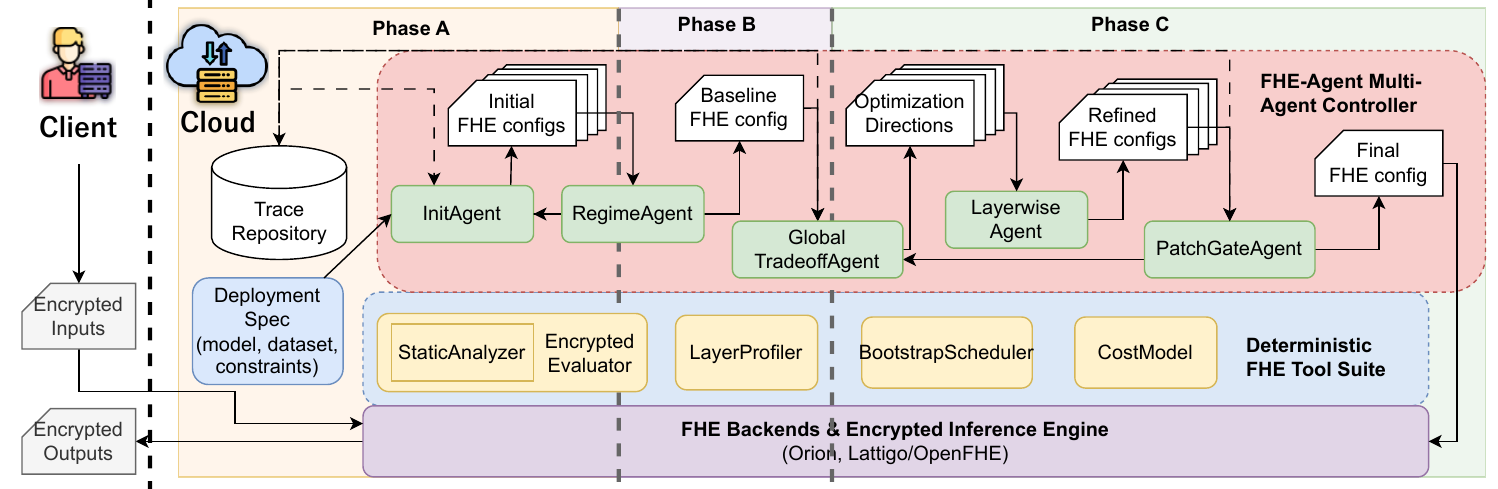}  
\vspace{-1em}
    \caption{
\textbf{Overview of FHE-Agent. A multi-agent controller (top) runs on the cloud and orchestrates a deterministic FHE tool suite (middle) on top of CKKS backends (bottom). Phase A performs simulation-only structure search, Phase B calibrates the cost model with a few encrypted runs, and Phase C applies admitted refinements under a global FHE budget. The trace repository stores configurations, metrics, and decisions for reuse across deployments, while the client only uploads encrypted inputs and receives encrypted outputs.}
}
    \label{fig:overview}  
\end{figure*}

Figure~\ref{fig:overview} illustrates the architecture of \sysname.
The system is composed of three functional layers:
(i) a deterministic \emph{FHE tool suite} that abstracts compiler-level signals and CKKS constraints;
(ii) a \emph{multi-agent controller} powered by a large language model (LLM) that orchestrates the search; and
(iii) a \emph{multi-fidelity evaluation backend} supported by a persistent trace repository.
To manage the high cost of encrypted evaluation, \sysname organizes the design space exploration into a three-phase workflow (Phase~A/B/C) that progressively increases fidelity while pruning the candidate space.

The tool suite encapsulates FHE-specific complexities (e.g. multiplicative depth analysis, security estimation, and packing logistics) ,behind stable, backend-agnostic interfaces.
Crucially, the agents are designed as \emph{decision-makers}, not calculators; they never perform numerical operations directly.
Instead, they invoke tools, analyze structured summaries, and select discrete \emph{optimization directions} (e.g., reducing the modulus chain length, switching the packing layout of a bottleneck layer, or relaxing bootstrap spacing).
The backend executes these candidates under a spectrum of evaluation modes ranging from static checks and cleartext simulation to fully encrypted inference, and logs all execution traces to the trace repository for few-shot context retrieval, debugging, and reuse as summarized exemplars when initializing new runs on related models.
Concretely, Section~\ref{subsec:tools} describes the FHE-aware tools, Section~\ref{subsec:agents} the multi-agent controller, Section~\ref{sec:layerwise} the layerwise profiling and constraints, and Section~\ref{sec:workflow} the three-phase optimization workflow.

\subsection{FHE Tool Suite and Backend Capabilities}
\label{subsec:tools}

We factor the FHE backend into a set of deterministic tools with clean input--output contracts.
This design ensures that the signals consumed by the agents are stable, interpretable, and decoupled from the underlying compiler implementation (e.g., Orion / Fhelipe).

\paragraph{StaticAnalyzer.}
The \textsf{StaticAnalyzer} validates the structural correctness of an FHE configuration $\mathcal{C}$ without running inference.
It verifies that the multiplicative depth required by linear layers and activation polynomials fits within the provided modulus chain, checks consistency in scale scheduling, and estimates the lattice security level $\mathrm{SecBits}(\mathcal{C})$ based on the CKKS parameter set $(\log N, \log Q, \sigma)$ using standard estimators~\cite{cheon2017ckks}.
It returns a tuple $(\mathrm{depth\_ok}, \mathrm{SecBits}(\mathcal{C}), \texttt{reasons})$, where \texttt{reasons} provides interpretable feedback on rejection (e.g., depth overflow or security violations).

\paragraph{LayerProfiler.}
Given an FHE configuration $\mathcal{C}$, a compiled model graph, and a calibration batch, the \textsf{LayerProfiler} executes the pipeline in \texttt{CLEAR\_ONLY} mode.
In this mode, ciphertexts are emulated via floating-point tensors to track value ranges, while the exact CKKS metadata (scale management, rotation keys, slot mapping) is fully preserved.
For each layer $i$, it aggregates structural, performance, and numerical statistics into a compact profile
\begin{equation}
  \label{eq:layer_profile}
  \mathsf{prof}_i = (\mathbf{s}_i, \mathbf{p}_i, \mathbf{n}_i),
\end{equation}
where $\mathbf{s}_i$ encodes layer type and tensor shapes,
$\mathbf{p}_i$ includes timing-related quantities such as the estimated runtime fraction, slot utilization, primitive counts (multiplications, rotations, bootstraps), and an approximate memory cost derived from active ciphertexts,
and $\mathbf{n}_i$ summarizes effective precision, layerwise approximation error, and remaining noise margin.
For non-linear or fused activation blocks, $\mathbf{n}_i$ also stores the degree and approximation error of the CKKS activation polynomial chosen by the backend; \sysname only reads these values and never modifies the polynomials themselves.
The remaining noise margin in bits is estimated by replaying the rescaling schedule in \texttt{CLEAR\_ONLY} mode using the backend's built-in noise simulator, so it can be interpreted as a conservative proxy for how close a layer is to exhausting its budget.

\paragraph{BootstrapScheduler.}
The \textsf{BootstrapScheduler} interfaces with the backend's resource allocator to manage depth consumption.
It computes the cumulative usage of the modulus chain and noise budget along all paths, generating a valid \emph{bootstrapping schedule} that partitions the graph into segments.
Crucially, it outputs a \emph{depth/noise mask} flagging layers that are critical to the budget.
This mask acts as a constraint for downstream agents, explicitly forbidding any modifications that would increase the multiplicative depth of these critical layers.

\paragraph{CostModel.}
The \textsf{CostModel} provides a rapid latency estimation $\mathsf{cost}(\mathcal{C})$ for an FHE configuration $\mathcal{C}$ by aggregating primitive counts from \textsf{LayerProfiler}:
\begin{equation}
  \label{eq:costmodel}
  \mathsf{cost}(\mathcal{C})
    = \sum_i \big(
      \alpha \cdot \texttt{mul}_i
    + \beta  \cdot \texttt{rot}_i
    + \gamma \cdot \texttt{boot}_i
    + \delta \cdot \texttt{mem\_cost}_i
    \big).
\end{equation}
The coefficients $(\alpha,\beta,\gamma,\delta)$ represent the unit costs of multiplications, key-switching/rotations, bootstrapping, and the overhead of handling wide layers with many active ciphertexts.
These coefficients are initialized via microbenchmarks and calibrated in Phase~B using encrypted runs (Section~\ref{sec:workflow}); the resulting model is then kept fixed during the subsequent refinement phase.

\paragraph{EncryptedEvaluator.}
We expose a unified API \texttt{run\_trial(config, eval\_mode)} that maps an FHE configuration \texttt{config} and an evaluation mode \texttt{eval\_mode} to a structured set of metrics.
The \texttt{eval\_mode} argument is an enum with four fidelity levels:
\texttt{STATIC\_ONLY} (checks logical validity using \textsf{StaticAnalyzer}),
\texttt{CLEAR\_ONLY} (profiles performance and precision via \textsf{LayerProfiler}, \textsf{BootstrapScheduler} and \textsf{CostModel} in cleartext simulation),
\texttt{FHE\_LIGHT} (runs encrypted inference on a mini-batch for calibration),
and \texttt{FHE\_FULL} (validates final accuracy and latency on the full set).
Here \texttt{config} denotes a complete FHE configuration, including the CKKS parameter set, scale schedule, bootstrapping plan, and packing/layout.
This multi-fidelity design allows the controller to reserve expensive encrypted cycles for only the most promising candidates.

\subsection{Multi-Agent Controller}
\label{subsec:agents}

\sysname employs a hierarchy of specialized LLM agents to navigate the configuration space.
This decomposition mirrors a human expert workflow: high-level architectural decisions are separated from low-level parameter tuning.

\paragraph{InitAgent and RegimeAgent (Exploration).}
The \textbf{InitAgent} performs cold-start exploration.
It combines the model summary (topology, tensor shapes) with user constraints (latency/accuracy budgets) and backend templates (e.g., ``high-precision'' vs.\ ``aggressive-packing'') to propose a diverse set of initial FHE configurations $\{\mathcal{C}^{(k)}\}$.
These regimes differ in the CKKS parameter set (e.g., ring dimension $\log N$ and total modulus size $\log Q$) and in global packing strategies.
The \textbf{RegimeAgent} then filters these candidates.
Using fast feedback from \texttt{STATIC\_ONLY} and \texttt{CLEAR\_ONLY}, it prunes infeasible regimes and selects one or two baseline configurations for detailed optimization.

\paragraph{GlobalTradeoffAgent and LayerwiseAgent (Optimization).}
Once a baseline FHE configuration $\mathcal{C}_{\mathrm{base}}$ is established, the \textbf{GlobalTradeoffAgent} proposes global strategy updates.
It analyzes global profiles and the bootstrapping schedule to suggest high-level patches, such as tightening the scale schedule, adjusting bootstrap spacing, or identifying a set of bottleneck layers $\mathcal{I}$ for targeted optimization.
In each iteration, the GlobalTradeoffAgent first selects this small set $\mathcal{I}$ and associated directional hints; it then delegates the fine-grained tuning of these bottlenecks to the \textbf{LayerwiseAgent}.
Restricted to the selected layers $i \in \mathcal{I}$, the LayerwiseAgent overrides local packing parameters (e.g., \texttt{embedding\_method}, \texttt{bsgs\_gap}) to optimize rotation counts and slot utilization.
It is not allowed to modify the CKKS parameter set or other scheme-level choices such as $(\log N, \log Q)$, nor the global scale schedule, and it must obey the \textsf{BootstrapScheduler}'s depth mask, ensuring local changes do not violate global depth constraints.

\paragraph{Discrete directions and patch space.}
To keep the search safe and interpretable, agents do not freely edit configuration files.
Instead, \sysname exposes a small vocabulary of \emph{optimization directions}, such as ``shorten tail of modulus chain,'' ``relax global scale by one step,'' ``switch a convolution from square to hybrid packing,'' or ``increase bootstrap interval by one layer.''
Each direction is compiled by the orchestrator into a structured patch on the current FHE configuration $\mathcal{C}$, affecting either a handful of global choices (e.g., CKKS parameter set or scale schedule) or a small set of layer-local fields.
This design constrains the patch space to changes that are known to be syntactically valid and semantically meaningful for the backend compiler, and it makes the agent’s decisions easier to audit and reuse.
\paragraph{PatchGateAgent and trace-guided admission control.}
To minimize wasted computation, the \textbf{PatchGateAgent} acts as a gatekeeper.
It simulates the batch of candidates proposed by the optimizer agents using \texttt{STATIC\_ONLY} and \texttt{CLEAR\_ONLY}.
A candidate configuration is admitted for encrypted evaluation (with \texttt{FHE\_LIGHT}) \emph{if and only if} it:
(i) passes all static security and depth checks,
(ii) satisfies precision gates (layerwise MAE and effective bits), and
(iii) shows the highest predicted latency gain according to the \textsf{CostModel}.
Admitted trials, together with their summaries and outcomes, are appended to the trace repository.
This admission control mechanism ensures that \sysname strictly bounds the number of expensive FHE executions while also accumulating a reusable history for future runs on similar architectures.

\vspace{-5pt}
\subsection{Layerwise Profiling and Constraints}
\label{sec:layerwise}

CKKS performance is often dictated by a few ``outlier'' layers (e.g., convolutions with awkward shapes).
From the profiles in~\eqref{eq:layer_profile}, \sysname derives a bottleneck score

\vspace{-.5em}
\begin{equation}
  \label{eq:bottleneck_score}
  \mathsf{score}_i
  = w_1 r_i + w_2 (1 - u_i) + w_3 \rho_i + w_4 z_i,
\end{equation}

where $r_i$ is the estimated runtime fraction of layer $i$, $u_i \in [0,1]$ is its slot utilization, $\rho_i$ is a normalized rotation count, and $z_i \in \{0,1\}$ indicates whether the remaining noise margin falls below a threshold.
In our implementation, these quantities are derived from the performance and numerical components of $\mathsf{prof}_i$ (e.g., $r_i$ from the runtime fraction field and $u_i$ from slot utilization).
Only the top-$K$ layers by $\mathsf{score}_i$ are exposed to the LayerwiseAgent as potential optimization targets, keeping the search focused on a small set of impactful bottlenecks rather than all layers.

To reflect the way human experts separate global parameter choices from local packing tweaks, we also view an FHE configuration as a hierarchical object
\begin{equation}
\footnotesize
  \label{eq:hier_config}
  \mathcal{C}
  = \big(\mathcal{C}_{\mathrm{global}}, \{\mathcal{C}^{(i)}_{\mathrm{local}}\}_i\big),
\end{equation}
where $\mathcal{C}_{\mathrm{global}}$ contains the CKKS parameter set $(N, Q, \sigma)$ and other global choices such as scale schedules and backend options, and each $\mathcal{C}^{(i)}_{\mathrm{local}}$ contains layer-specific packing and embedding choices.
\sysname enforces that only the exploration agents (InitAgent and GlobalTradeoffAgent) may modify $\mathcal{C}_{\mathrm{global}}$, while the LayerwiseAgent is restricted to a small subset of $\mathcal{C}^{(i)}_{\mathrm{local}}$ for bottleneck layers and must respect the depth/noise mask from \textsf{BootstrapScheduler}.
This hierarchical structure reduces the effective search space and prevents local tweaks from accidentally violating global FHE constraints anchored by the CKKS parameter set.

To ensure numerical stability, we enforce \emph{feasibility gates}.
Unlike approaches that re-train the model, \sysname treats the backend's activation polynomials as fixed, but imposes hard constraints on the resulting layerwise approximation error and effective precision.
Combined with the global security and depth checks from \textsf{StaticAnalyzer}, these gates ensure that any FHE configuration that degrades these metrics beyond a threshold (e.g., due to overly aggressive scaling) is rejected during the \texttt{CLEAR\_ONLY} simulation, preventing the system from wasting cycles on numerically unstable designs.

\subsection{Three-Phase Optimization Workflow}
\label{sec:workflow}

The optimization process is structured to maximize information gain per encrypted trial.

\noindent\textbf{Phase A: Structure Search (Simulation-Only).}
The goal is to identify feasible structural regimes.
The InitAgent proposes candidate FHE configurations which are evaluated strictly in \texttt{STATIC\_ONLY} and \texttt{CLEAR\_ONLY} modes via \textsf{StaticAnalyzer}, \textsf{LayerProfiler}, \textsf{BootstrapScheduler}, and \textsf{CostModel}.
Candidates violating security or depth gates are immediately discarded.
Surviving regimes are ranked by their proxy latency (Eq.~\ref{eq:costmodel}), slot utilization, and precision scores, and only a small number advance to calibration.

\noindent\textbf{Phase B: Calibration and Selection.}
We bridge the reality gap between simulation and execution.
For the top regimes from Phase~A, the backend runs \texttt{FHE\_LIGHT} on a small validation subset using the \textsf{EncryptedEvaluator}.
These runs yield ground-truth latency and noise measurements, which are used to regress the coefficients of the \textsf{CostModel}.
The RegimeAgent then picks the best-performing configuration as the calibrated baseline FHE configuration $\mathcal{C}_{\mathrm{base}}$.
In our current prototype, cost-model calibration is performed once at the end of Phase~B, and the resulting coefficients are kept fixed during Phase~C.

\noindent\textbf{Phase C: Admitted Refinement.}
Starting from $\mathcal{C}_{\mathrm{base}}$, the agents iteratively refine the design.
In each step, the GlobalTradeoffAgent and LayerwiseAgent propose patches drawn from the discrete direction vocabulary.
The PatchGateAgent simulates them with \texttt{STATIC\_ONLY} and \texttt{CLEAR\_ONLY} and admits at most one candidate per iteration for \texttt{FHE\_LIGHT} evaluation, updating the trial history while relying on the fixed calibrated cost model.
A global budget on encrypted trials limits the number of iterations; within this budget, the controller prefers directions that yield the largest predicted latency reduction while keeping all feasibility gates satisfied.
For large models (e.g., ResNet), where even light evaluation is costly, \texttt{FHE\_FULL} is reserved strictly for the final verification of the converged solution.
This workflow ensures that \sysname scales effectively from small MLPs to deep CNNs while keeping the number of fully encrypted runs manageable.

%% file: Sections/05_evaluation.tex
\vspace{-5pt}
\section{Evaluation}

\subsection{Experimental Setup and Metrics}
\label{sec:setup}
\paragraph{Hardware and backends.}
All experiments run on a dual-socket server with two AMD EPYC~9454 processors
(48 cores per socket, 192 hardware threads) and 1.5\,TiB of RAM.
We instantiate \sysname on top of the Orion FHE compiler~\cite{ebel2025orion}
targeting the Lattigo v5.0.2 CKKS backend~\cite{mouchet2021lattigo};
both the Orion examples and \sysname-generated configurations are compiled
and executed through this Orion+Lattigo stack under single-threaded execution
so that we isolate the effect of different FHE configurations rather than
parallelism.
The multi-agent controller invokes a commercial LLM via API only to select
optimization directions, while all numerical reasoning is delegated to the
deterministic tools in Section~\ref{subsec:tools}.
We follow the training and preprocessing recipes of the corresponding Orion
examples (MNIST for MLP/LeNet/LoLa~\cite{lecun1998mnist} and CIFAR-10 for
AlexNet~\cite{alexnet10.5555/2999134.2999257}), and, when available, start from the
Orion reference FHE configuration for each model to fix the CKKS parameter set
and activation polynomials.

\paragraph{Metrics.}
For each FHE configuration $\mathcal{C}$ we report:
(i) encrypted-task accuracy and its gap to plaintext,
(ii) mean absolute error (MAE) between encrypted and cleartext outputs,
(iii) effective precision (in bits) estimated from the output noise margin,
(iv) FHE runtime as the end-to-end encrypted inference time per input on the
validation set (excluding key generation), and
(v) the estimated security level $\mathrm{SecBits}(\mathcal{C})$ derived from
the CKKS parameter set $(\log N, \log Q, \sigma)$ using our CKKS estimator
(Section~\ref{subsec:problem_formulation}).
When comparing Orion and \sysname, we fix the target security level and
highlight differences in latency, modulus-chain depth, and bootstrapping cost.

\begin{table}[t]
\centering
\small
\resizebox{\columnwidth}{!}{
\begin{tabular}{l l c c c c}
\toprule
Model & Method &
Precision (bits)$\uparrow$ &
MAE$\downarrow$ &
FHE time [s]$\downarrow$ &
Sec. [bits]$\uparrow$ \\
\midrule
\multirow{2}{*}{MLP} 
  & Naive search 
    & 17.37 & $5.89\times 10^{-6}$ & 1.31  & $\ge 128$ \\
  & \sysname 
    & 24.82 & $\approx 0$          & 0.91  & $\ge 128$ \\
\midrule
\multirow{2}{*}{LeNet} 
  & Naive search 
    & 23.07 & $1.13\times 10^{-7}$ & 9.08  & $\ge 128$ \\
  & \sysname 
    & 22.42 & $1.79\times 10^{-7}$ & 3.19  & $\ge 128$ \\
\midrule
\multirow{2}{*}{LoLA} 
  & Naive search 
    & 19.54 & $1.31\times 10^{-6}$ & 2.10  & $\ge 128$ \\
  & \sysname 
    & 21.16 & $\approx 0$          & 0.79  & $\ge 128$ \\
\midrule
\multirow{2}{*}{AlexNet} 
  & Naive search 
    & N/A   & N/A                  & N/A   & N/A       \\
  & \sysname 
    & 21.81 & $\approx 0$          & 262.5 & $\ge 128$ \\
\bottomrule
\end{tabular}
}
\caption{Best FHE configurations found by direct LLM configuration search
(\emph{Naive search}) and by \sysname under a fixed 128-bit security target.}
\label{tab:summary-naive-vs-agent}
\vspace{-3em}
\end{table}

\subsection{Overall results: agent vs naive search}
We first evaluate a naive ``one-shot'' LLM baseline.
For each model, we prompt the same commercial LLM used inside \sysname
\emph{ten} times to directly generate a full Orion configuration, given the
plaintext model description, dataset, and high-level constraints (e.g.,
``$\ge 128$-bit security, small accuracy loss'').
Each proposed configuration is passed through our tool suite; if it satisfies
all feasibility gates, we run encrypted inference and record its metrics.
The ``Naive search'' row in Table~\ref{tab:summary-naive-vs-agent} reports,
for each model, the best feasible configuration among these ten one-shot
trials (or N/A if no suggestion passes all checks).
This baseline reflects how well the LLM can serve as a stand-alone FHE
configuration engine when it must emit the entire Orion configuration in one
shot.

Using the same underlying LLM, \sysname instead constrains the model to
choosing discrete optimization directions on top of the deterministic tools
and multi-fidelity backend from Section~\ref{sec:system}.
The ``\sysname'' row in Table~\ref{tab:summary-naive-vs-agent} reports the
best configuration produced by a single run of our three-phase workflow for
each model, under the same CKKS backend and security target.
Across MLP, LeNet, and LoLA, the agentic workflow consistently finds
configurations that satisfy our gates and either match or improve the naive
baseline's numerical quality while reducing FHE runtime: for MLP, \sysname
improves precision from 17.37 to 24.82 bits and reduces runtime from
1.31\,s to 0.91\,s; for LeNet and LoLA, it achieves similar or higher
precision with roughly $3\times$ faster encrypted inference.
For AlexNet, none of the ten one-shot LLM configurations survive the feasibility checks, so the naive baseline has no feasible point; in contrast, \sysname is able to drive the same backend to a 128-bit secure configuration with 21.81 bits of effective precision, near-zero MAE, and a finite FHE runtime of 262.5\,s, demonstrating that the agentic workflow can still recover valid CKKS settings even for deeper CNNs where direct prompting fails completely.

Because most candidate directions are filtered in
\texttt{STATIC\_ONLY}/ \texttt{CLEAR\_ONLY} modes, \sysname also requires
fewer fully encrypted trials than the naive baseline, which evaluates all
ten one-shot suggestions under FHE.
Overall, these aggregate results indicate that a tool-guided, multi-agent
controller can turn the same base LLM into a more stable and efficient FHE
configuration engine.
In the next subsection, we zoom in on LeNet to show how layerwise profiling
and feasibility gates drive this behavior under a fixed CKKS parameter set.



\begin{table}[t]
  \centering
  \small
\resizebox{\columnwidth}{!}{
  \begin{tabular}{lcccc}
    \toprule
    & \multicolumn{4}{c}{Trial (fixed CKKS security, $\log N = 15$)} \\
    Global metrics & 0 & 1 & 2 & 3 \\
    \midrule
    Total runtime [s]      & 7.89 & 6.25 & 5.04 & 8.51 \\
    MAE                    & $3.0\times10^{-4}$ & $2.9\times10^{-2}$ & $2.9\times10^{-2}$ & $1.6\times10^{-3}$ \\
    Precision [bits]       & 11.63 & 5.12 & 5.12 & 9.27 \\
    \# rotations           & 8 & 7 & 7 & 8 \\
    \# multiplications     & 4 & 4 & 4 & 4 \\
    \# bootstraps          & 0 & 0 & 0 & 0 \\
    Act.\ degree (conv1)   & 31 & 31 & 31 & 15 \\
    \midrule
    Layer & \multicolumn{4}{c}{\textit{Per-layer runtime [s (share of total)]}} \\
    conv1 & 2.483 (32.0\%) & 2.561 (41.3\%) & 2.555 (56.4\%) & 2.516 (29.9\%) \\
    conv2 & 4.006 (51.6\%) & 2.794 (45.1\%) & 1.610 (32.1\%) & 4.304 (51.1\%) \\
    fc1   & 0.906 (11.7\%) & 0.548 (8.8\%)  & 0.450 (9.0\%)  & 1.016 (12.1\%) \\
    fc2   & 0.368 (4.7\%)  & 0.298 (4.8\%)  & 0.399 (8.0\%)  & 0.584 (6.9\%) \\
    \bottomrule
  \end{tabular}
  }
\caption{LeNet case study with fixed CKKS parameters
($\log N = 15$, 256-bit security). Trial~0 is the Phase~A/B
agent configuration; trials~1--3 are later agent refinements.}
  \label{tab:lenet_case}
\vspace{-3em}
\end{table}

\vspace{-5pt}
\subsection{Case study: LeNet under fixed CKKS.}
\label{sec:lenet-case-study}

To illustrate how \sysname uses layerwise profiling and encrypted feedback to guide search, we present a LeNet case study based on Orion's MNIST example.
To focus on the agent’s behavior under a fixed CKKS regime, we fix the CKKS parameter set to $\log N = 15$ with a 256-bit security target and reuse the backend's activation polynomials.
Across all trials in Table~\ref{tab:lenet_case}, the CKKS parameter set and total modulus size are therefore held fixed; the controller is only allowed to adjust layerwise packing and activation degrees based on the tool suite.

Under these constraints, Phases~A and~B run once to identify an initial feasible configuration, which we denote as Trial~0 in Table~\ref{tab:lenet_case}.
Starting from Orion's reference configuration (used only to fix the CKKS parameter set and activation polynomials), the InitAgent and RegimeAgent search over packing choices and scale schedules within this fixed regime and select a configuration that passes all feasibility gates, with MAE $3.0\times 10^{-4}$ and 11.63 bits of effective precision.
The \textsf{LayerProfiler} reports that the second convolution (\texttt{conv2}) dominates the runtime (4.006\,s, 51.6\% of the total), while \texttt{conv1}, \texttt{fc1}, and \texttt{fc2} contribute 32.0\%, 11.7\%, and 4.7\%, respectively, for a total FHE runtime of 7.89\,s per image.
Because the MAE and precision gates are easily satisfied, the InitAgent and RegimeAgent keep this FHE configuration as $\mathcal{C}_{\mathrm{base}}$ and hand control to the GlobalTradeoffAgent and LayerwiseAgent, which then perform Phase~C layerwise exploration to search for packing-level optimizations that reduce latency without changing the CKKS parameter set.

We report four fully encrypted trials (Trial~0--3); an additional \texttt{CLEAR\_ONLY} profiling run is omitted because it does not involve FHE execution.
Trial~1 is the first Phase~C refinement after \texttt{conv2} has been identified as the primary bottleneck.
Guided by the layerwise profile, the LayerwiseAgent applies a more aggressive packing override on \texttt{conv2}.
This reduces its runtime from 4.006\,s to 2.794\,s and lowers its share from 51.6\% to 45.1\%, bringing the overall FHE runtime down to 6.25\,s.
However, the feasibility gates detect that the output MAE rises to $2.9\times 10^{-2}$ and the effective precision drops to 5.12 bits, so Trial~1 violates the MAE and precision constraints.
Trial~2 continues to optimize the same regime: \texttt{conv2} runtime is further reduced to 1.610\,s (32.1\% of total) and the overall runtime reaches 5.04\,s, but MAE and precision remain essentially unchanged at $2.9\times 10^{-2}$ and 5.12 bits, so this trial is also rejected.

At this point, the PatchGateAgent prevents \sysname from accepting the faster but numerically unstable configurations and steers the search back toward the feasible region.
Using updated layerwise profiles and noise margins, the GlobalTradeoffAgent flags the first convolution (\texttt{conv1}) as a candidate where a lower-degree activation polynomial and reduced parallelism can recover precision without exhausting the noise budget.
The LayerwiseAgent then applies two local overrides to \texttt{conv1}: it lowers the activation degree from 31 to 15 and caps the parallelism via \texttt{max\_parallel\_blocks}~=~2.
Trial~3 evaluates this patch and restores MAE to $1.6\times 10^{-3}$ and effective precision to 9.27 bits, comfortably within the feasibility gates, while keeping the CKKS security level unchanged.
As Table~\ref{tab:lenet_case} shows, the number of ciphertext multiplications and bootstraps is identical across trials (4 and 0, respectively), and the rotation count only fluctuates between 7 and 8, confirming that the search operates within a fixed depth and bootstrapping regime.

Overall, this LeNet case study shows how \sysname realizes a controlled, tool-informed exploration loop under a fixed CKKS parameter set.
The \textsf{StaticAnalyzer} and \textsf{LayerProfiler} identify bottlenecks, the GlobalTradeoffAgent and LayerwiseAgent propose layer-specific directions (e.g., more aggressive packing on \texttt{conv2}, activation-degree and parallelism changes on \texttt{conv1}), and the PatchGateAgent enforces feasibility gates and a strict FHE budget.
In this example, the controller reaches a constraint-satisfying configuration after only four fully encrypted trials, with the rest of the reasoning performed in simulation.

%% file: Sections/06_conclusion.tex
\section{Conclusion}
Configuring CKKS-based encrypted inference is a key obstacle to
practical FHE deployment: small changes to parameters or packing can
break security or precision or inflate latency by orders of magnitude.
We framed this as a resource-constrained configuration search problem
and proposed \textbf{FHE-Agent}, which combines an LLM-based multi-agent
controller with a deterministic FHE tool suite and a multi-fidelity
evaluation backend.

By exposing static analyzers, layerwise profilers, and cost models as
tools and restricting the agents to safe, discrete configuration
directions, FHE-Agent can aggressively prune the search space using
static and cleartext feedback and reserve fully encrypted runs for a few
promising candidates.
Our prototype on Orion+Lattigo shows that this workflow automatically
discovers high-quality configurations across MLP, LeNet, LoLA, and
AlexNet: compared to a naïve one-shot LLM search, FHE-Agent achieves
similar or better precision at noticeably lower encrypted runtime, and
recovers feasible 128-bit–secure configurations even for architectures
where direct configuration generation by the LLM fails.

Looking ahead, we see several directions for future work.
On the systems side, FHE-Agent could be extended to additional FHE
libraries and to GPU-accelerated backends, as well as to other
approximate schemes beyond CKKS.
On the algorithmic side, integrating more accurate security estimators,
richer noise-tracking models, and more principled policies for direction
selection may further reduce the number of encrypted trials.
Ultimately, we hope that agentic orchestration of compiler tool suites
can make encrypted MLaaS configuration routine, shifting FHE deployment
from expert-driven tuning to automated, auditable workflows.